\documentclass[twocolumn, prd, aps, nofootinbib, showpacs,showkeys, %preprint
]{revtex4-1}

\pdfoutput=1
\usepackage{epsf,epsfig,subfigure,graphicx,amsmath,amssymb}
\usepackage{color}
\usepackage{hyperref}

\def\lsim{\mathrel{\rlap{\lower4pt\hbox{\hskip1pt$\sim$}}
    \raise1pt\hbox{$<$}}}         %less than or approx. symbol
\def\gsim{\mathrel{\rlap{\lower4pt\hbox{\hskip1pt$\sim$}}
    \raise1pt\hbox{$>$}}}

\newcommand{\be}{\begin{eqnarray}}
\newcommand{\ee}{\end{eqnarray}}
\newcommand{\bea}{\begin{eqnarray}}
\newcommand{\eea}{\end{eqnarray}}

\begin{document}

%\begin{flushright}
%{\tt .........}
%\end{flushright}

\title{A testable scenario of WIMPZILLA with Dark Radiation}

\author{Jong-Chul Park %\footnote{log1079@gmail.com}
and Seong Chan Park }%\footnote{s.park@skku.edu}}

\email{s.park@skku.edu}

\address{Department of Physics, Sungkyunkwan University, Suwon 440-746, Korea}

%\vspace{1.0cm}
\begin{abstract}
As the electromagnetic gauge symmetry makes the electron stable, a new abelian gauge symmetry may be responsible for the stability of superheavy dark matter.
The gauge boson associated with the new gauge symmetry naturally plays the role of dark radiation and contributes to the effective number of  `neutrino species', which has been recently measured by Planck. We estimate the contribution of dark radiation from the radiative decay of a scalar particle induced by the WIMPZILLA  in the loop. The scalar particle  may affect the invisible decay of the Higgs boson by the Higgs portal type coupling.
\end{abstract}

 %\pacs{95.35.+d, 95.85.Ry, 98.70.Rz, 12.60.Cn}

 %95.35.+d DM
 %98.62.Gq galactic halos
 %98.70.Rz gamma ray sources
 %95.85.Ry neutrino muon and other elem. particles,
 %11.30.Ly: other internal and higher symmetries
 %12.60.Cn: extension of EW sector
 %12.90.+b: Miscellaneous models
 %14.70.Pw: Other gauge bosons


\maketitle

%%%%%%%%%%%%%%%%%%%%%%%%%%%%%%%%%%%%%%%%%%%%%%%%%%%%%%%%%%%%%%%%
%%%%%%%%%%%%%%%%%%%%%%%%%%%%%%%%%%%%%%%%%%%%%%%%%%%%%%%%%%%%%%%%

\section{Introduction}

Having no firmly established evidence for weakly interacting massive particle (WIMP) at the LHC and direct detection experiments below a TeV, we are motivated to consider  dark matter in a heavier energy regime. If the dark matter particle is heavier than a TeV, however, some crucial difficulties arise as follows.
\begin{itemize}
\item {\it Overclosure of universe:} The annihilation cross-section, $\langle \sigma v \rangle \propto 1/m_{DM}^2$, becomes small and leads to the overclosure of the universe with a large density, $\Omega_{DM} >1$.
\item  {\it Instability of dark matter:} A heavy particle, in general, tends to be unstable if no symmetry principle forbids its decay.% fundamentally due to the uncertainty principle, $\Delta t \sim 1/\Delta E \sim 1/M_{DM}$.
\item {\it Lack of testability:} In currently on-going or future running experiments,  heavier dark matter well above a TeV range is hard to get tested \cite{Bertone:2004pz, Feng:2010gw}.
\end{itemize}

Rather surprisingly, the overclosure of universe can be resolved in the case of superheavy dark matter in a mass window $m_{DM} \approx 10^{12-14}$ GeV. For a reheating temperature about $10^9$ GeV, it has been shown that the superheavy dark matter, dubbed WIMPZILLA, provides nice fit to the observed dark matter abundance \cite{WIMPZILLA}.

However, the stability problem is worse for WIMPZILLA with its extremely high mass. A global symmetry, such as R-parity in supersymmetry models, does not work since in the presence of gravity, especially in the vicinity of black hole, all the stable particles (strings and branes) should be associated with gauge symmetries \cite{Banks:2010zn}. An interesting observation is that the electron is stable due to the exact electro-magnetic ${\rm U(1)}_{em}$ gauge symmetry in the standard model (SM). Here we consider the similar possibility that an abelian group ${\rm U(1)}_{H}$ is responsible for the stability of superheavy dark matter. The gauge symmetry may originate from a compact group such as ${\rm E}_6$ \cite{Hewett:1988xc}. Just like the photon with ${\rm U(1)}_{em}$, a new massless gauge boson is associated with the new gauge symmetry and the presence of new gauge boson opens a new window of testability of WIMPZILLA!

The new gauge boson does not directly interact with the SM sector since it is associatively introduced for dark matter. We thus regard the new gauge boson as hidden photon or dark radiation.  Dark radiation may have left the evidence of its presence in various circumstances in cosmological history of the universe. Indeed, recent observations show that there may exist non-standard model relativistic particles at the time of Big Bang nucleosynthesis (BBN) and also at the era of recombination shown in cosmic microwave background radiation (CMBR).

In the rest of this paper, we examine possible experimental tests of this WIMPZILLA associated dark radiation in BBN and CMBR \cite{Fischler:2010xz, Menestrina:2011mz, Choi:2012zna, GonzalezGarcia:2012yq, Hasenkamp:2012ii} and also collider physics \cite{Kim:2009qc} based on a model, which is shortly introduced in the next section. In the following section \ref{sec:DR}, we calculate the expected amount of dark radiation from the hidden sector scalar decay then discuss how the scalar particle can affect the Higgs boson decay patterns in section \ref{sec:higgs}. The summary is given in the last section.

\section{The Model}

The model includes two new hidden sector particles: $\psi$ and $\phi$. A Dirac fermion field, $\psi$, is charged under the extra abelian gauge symmetry, ${\rm U(1)}_H$, and identified with dark matter. A massive scalar particle, $\phi$, is neutral under the gauge symmetry as well as the standard model interactions but is responsible for the late time decay to dark radiation. The generic gauge invariant Lagrangian\footnote{The kinetic mixing term $\sim F_{\mu\nu}F_H^{\mu\nu}$ can be another source of communication between the hidden sector and the SM sector \cite{KineticMixing, KineticMixing2} if there exists a bi-charged particle of both sectors, which is absent in the current model. } is given by
\begin{eqnarray}\label{Interactions}
{\cal L}&& \supset {\cal L}_{SM}
+ \lambda_{\phi H} \phi^2 (H^\dagger H)
- \frac{1}{4}\, F_{H \mu\nu}F_H^{\mu\nu} \nonumber \\
&&- y_\psi \phi \overline{\psi}\psi + i
\overline{\psi}\gamma^\mu (\partial_\mu - i g_H A^H_\mu) \psi
- m_\psi \overline{\psi}\psi\,,
\end{eqnarray}
where the `Higgs portal' interaction with the coupling constant $\lambda_{\phi H}$ \cite{HiggsPortal} is allowed.

%
%%%%%%%%%%%%%%%%%%%%%%%%%%%%%%%%%%%%%%%%%%%%%%%%%%%%%%%%%%%%
\begin{figure}[t]
\begin{center}
\includegraphics[width=0.7\linewidth]{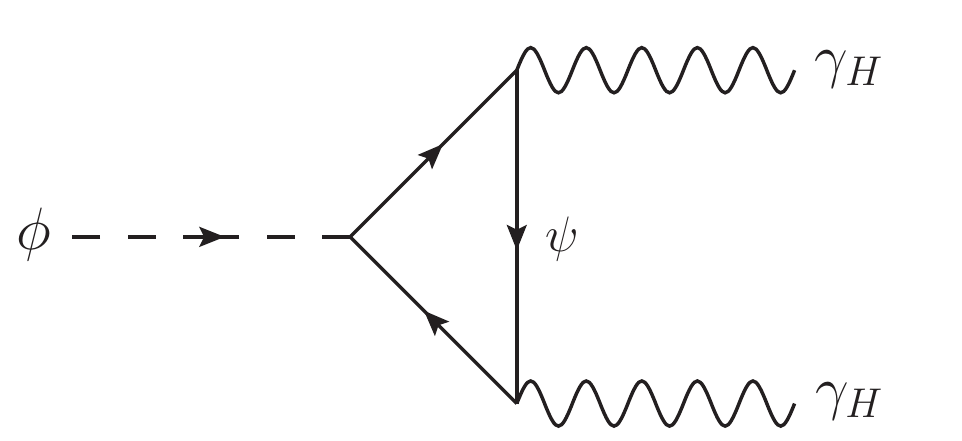}
\end{center}
\caption{A scalar particle ($\phi$)  decaying into two hidden photons through the WIMPZILLA ($\psi$) loop.}
\label{Fig_scalardecay}
\end{figure}
%%%%%%%%%%%%%%%%%%%%%%%%%%%%%%%%%%%%%%%%%%%%%%%%%%%%%%%%%%%%
%

Differently from the dark matter particle, which is protected by ${\rm U(1)}_H$, the scalar particle is unstable and  decays to two hidden photons  at one-loop level through the triangle diagram with the virtual $\psi$ in the loop as can be seen in
Fig.~\ref{Fig_scalardecay}. The decay width of $\phi$ is highly suppressed by the heavy mass of $\psi$:
\begin{eqnarray}
\frac{1}{\tau_{\phi \to \gamma_H\gamma_H}}\approx \frac{(\alpha_H y_\psi)^2}{144 \pi^3} \frac{m_\phi^3}{m_\psi^2}\,,
\end{eqnarray}
where the structure constant is $\alpha_H = g_H^2/4\pi$. %and the loop function is $F(\tau) =
%-2 \tau \left[1 + (1 - \tau)\arcsin^2(1/\sqrt{\tau})\right]$ with $\tau= 4m_\psi^2/m_\phi^2$. The loop function is well approximated by $-4/3$ at a large $\tau$ or heavy mass limit.
%For a given length of life time, the combination of coupling constants is written as
%\begin{eqnarray}
%\alpha_Hy_\psi
%=9.64\cdot10^{-2} \sqrt{\tfrac{100s}{\tau_\phi}} \left( \tfrac{m_\psi}{10^{13}\, {\rm GeV}} \right) \left( \tfrac{100\, {\rm GeV}}{m_\phi} \right)^{\tfrac{3}{2}}\,.
%\end{eqnarray}
%
With reasonable choice of parameters, the life time can lie in particularly interesting epochs:
\begin{itemize}
\item the BBN epoch, $t_{\rm BBN} \approx \mathcal{O}(0.1-1000){\rm s}$,
\item the CMB epoch, $t_{\rm CMB} \approx 3.8 \times 10^5 {\rm yr} \approx 1.2 \times 10^{13} {\rm s}$,
\end{itemize}
thus, in principle, we can test the idea of WIMPZILLA by observing dark radiation in these epochs.

\section{Observational bounds}

In this section, we consider dark radiation components seeing in CMB and BBN data and the possible contribution to the Higgs invisible decay at the LHC.

\subsection{dark radiation \label{sec:DR}}

The effective number of relativistic degrees of freedom, $N_{\rm eff}$, is recently observed by Planck~\cite{Planck2013}, WMAP 9 year~\cite{WMAP9} and also BBN ~\cite{BBN}:
\begin{eqnarray}
N^{\rm CMB}_{\rm eff} &=& 3.30 \pm 0.27 \quad \text{(Planck 2013)}, \nonumber \\
N^{\rm CMB}_{\rm eff} &=& 3.84 \pm 0.40 \quad \text{(WMAP9)}, \label{eq:neff}
 \\
N^{\rm BBN}_{\rm eff} &=& 3.71^{+0.47}_{-0.45}  \quad \quad \text{(BBN)}. \nonumber
\end{eqnarray}
Compared to the SM expectation, $N^{\rm SM}_{\rm eff} = 3.046$ \cite{NeffSM}, there exist some deviation in particular in WMAP9 and BBN results.

When $\phi$ never has dominated the expansion of the universe, but produced extra relativistic degrees of freedom by its decay \cite{Fischler:2010xz, Menestrina:2011mz, Choi:2012zna, GonzalezGarcia:2012yq, Hasenkamp:2012ii}, the extra contribution to the effective number of relativistic degree of freedom, $\Delta N_{\rm eff}$, is calculated with $Y_\phi (=n_\phi/s) m_\phi$ and $\tau_\phi$ by a simple formula
\cite{Scherrer:1987rr, Menestrina:2011mz}
\begin{eqnarray}\label{DR_decay}
\Delta N_{{\rm eff}, \phi} = 8.3 (Y_\phi m_\phi/{\rm MeV})(\tau_\phi/{\rm s})^{1/2}\,.
\end{eqnarray}

The primordial hidden photon $\gamma_H$ also contributes to $N_{\rm eff}$ but the effect is small:
\begin{eqnarray}
\Delta N_{{\rm eff}, \gamma_H} &=& \frac{2}{(7/8)\cdot 2} \left(\frac{11}{4}\right)^{4/3}
\left(\frac{g_{*S, 0}}{g_{*S, \gamma_H {\rm dec}}}\right)^{4/3} \nonumber \\
 &\simeq& 0.053\,,
\end{eqnarray}
where we have used $g_{*S, 0} = 3.91$ and $g_{*S, \gamma_H {\rm dec}} = 107.75$ for the total degrees of freedom associated with entropy at present and the decoupling time of $\gamma_H$.

Finally, we get the total $N_{\rm eff}$ with the SM contribution, the decay of $\phi$ and also the primordial $\gamma_H$:
\begin{eqnarray}
N_{\rm eff} = N^{\rm SM}_{\rm eff} +  \Delta N_{{\rm eff}, \phi} + \Delta N_{{\rm eff}, \gamma_H}\,,
\end{eqnarray}
which we should compare with the observational results in Eq. \eqref{eq:neff}, in particular, the Planck 2013 result.

\begin{widetext}
%aaa

%
%%%%%%%%%%%%%%%%%%%%%%%%%%%%%%%%%%%%%%%%%%%%%%%%%%%%%%%%%%%%
\begin{figure}[h]
\begin{center}
\includegraphics[width=0.32\linewidth]{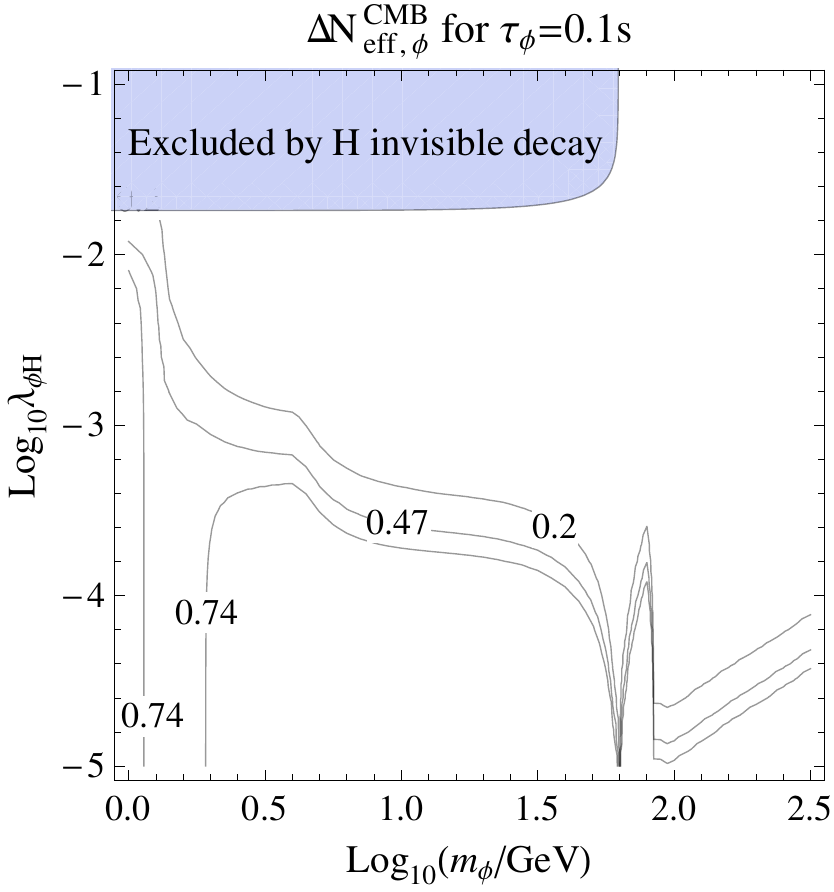}
\includegraphics[width=0.32\linewidth]{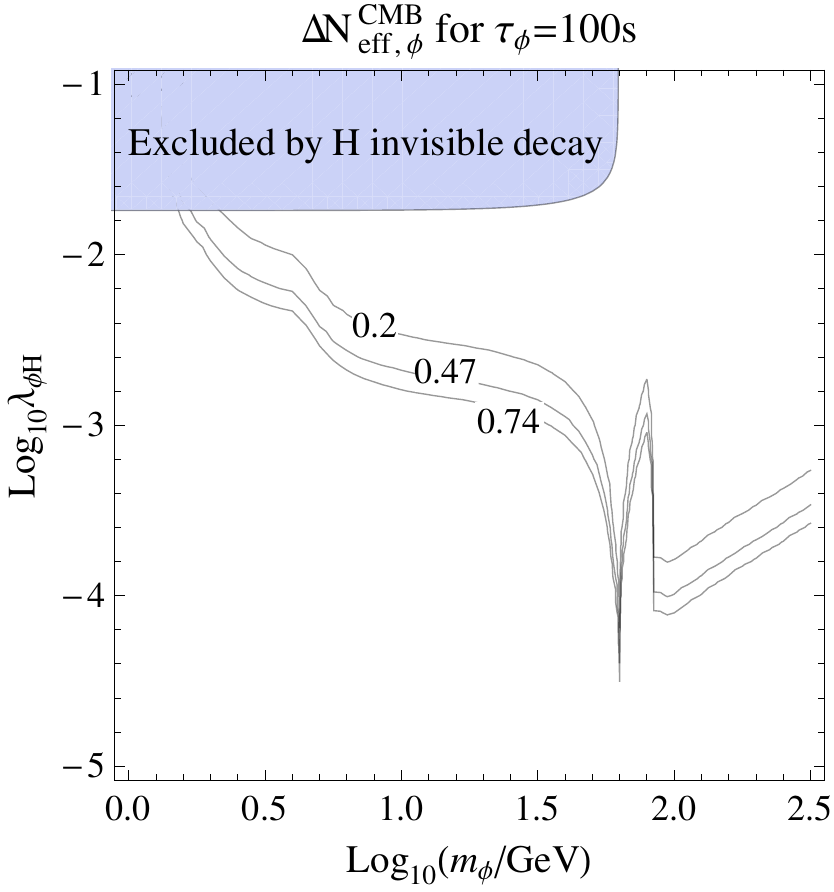}
\includegraphics[width=0.32\linewidth]{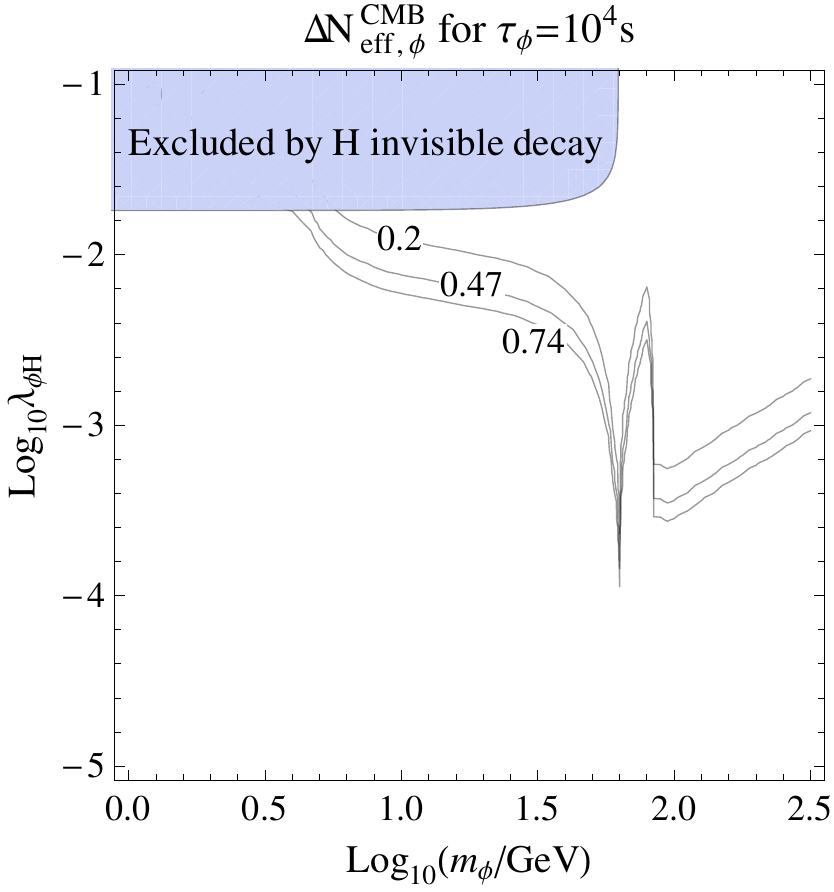}
\end{center}
\caption{Contour plots for $\Delta N_{{\rm eff}, \phi}^{\rm CMB}$ in the $m_\phi - \lambda_{\phi H}$ plane when the decay life time of $\phi$, $\tau_\phi$, is fixed as 0.1, 100 and $10^4$ s from left to right panels. The shaded region is excluded by the conservative invisible Higgs decay width limit \cite{HiggsFit}, ${\rm BR}_{\rm inv} < 0.28$ at $95\%$ C.L. allowing non-standard values for $h \rightarrow \gamma\gamma$ and $h \rightarrow gg$ modes.
The left-lower regions below solid lines of 0.47 and 0.74 are constrained by the effective number of relativistic degrees of freedom limit from the Planck observation~\cite{Planck2013} at $1 \sigma$ and $2 \sigma$ level, respectively.}
\label{fig:cons}
\end{figure}
%%%%%%%%%%%%%%%%%%%%%%%%%%%%%%%%%%%%%%%%%%%%%%%%%%%%%%%%%%%%
%
\end{widetext}

In Fig.~\ref{fig:cons}, we plotted $\Delta N_{{\rm eff}, \phi}$ with the Higgs portal coupling $\lambda_{\phi H}\in (10^{-5}, 10^{-1})$ with respect to $m_\phi$ in $(10^0, 10^{2.5})$ GeV  when the life time of $\phi$ is given as $10^{-1}s$ (left), $10^2s$ (middle) and $10^4s$ (right), respectively. The left-lower regions below solid lines of 0.47 and 0.74 are constrained by the effective number of relativistic degrees of freedom limit from the Planck observation \cite{Planck2013} at $1\sigma$ and $2\sigma$ level, respectively.

\subsection{Invisible decay of the Higgs \label{sec:higgs}}

Recently, ATLAS and CMS reported the discovery of a SM Higgs-like new particle with the mass of about 125 GeV~\cite{ATLAS, CMS}. Having non-vanishing `Higgs-portal' interaction, which induces communication between the hidden and visible sectors, the decay of the Higgs boson to the hidden sector scalar particles is available when $2 m_\phi < m_h$. When the width to the invisible particles, i.e. a pair of $\phi$'s, is sizable, the decay process can be detected by the LHC experiment. A global fit analysis to all the Higgs search data provides a limit on the invisible Higgs decay width: ${\rm BR}_{\rm inv} < 0.19$ at $95\%$ C.L. assuming the SM decay rates for all the visible Higgs decay modes and  ${\rm BR}_{\rm inv} < 0.28$ at $95\%$ C.L. allowing non-standard values for $h \rightarrow \gamma\gamma$ and $h \rightarrow gg$ modes \cite{HiggsFit}.
We obtained a constraint on the Higgs portal coupling $\lambda_{\phi H}$ from the invisible Higgs decay limit. In this analysis, we used ${\rm BR}_{\rm inv} < 0.28$ as a conservative bound and plotted in Fig.~\ref{fig:cons}. The upper left shaded region is excluded.

\section{Conclusion}

Superheavy dark matter with $m_{DM} \approx 10^{12-14}$ GeV, dubbed WIMPZILLA, can satisfy the observed dark matter abundance but the stability of WIMPZILLA requires an explanation.  In this paper, a new gauge symmetry ${\rm U(1)}_H$ is introduced to stabilize WIMPZILLA just like the electromagnetic gauge symmetry to the electron. The gauge boson associated with the new gauge symmetry is naturally interpreted as dark radiation and provides possible observational consequences in CMBR and BBN data.
The abundance of dark radiation is determined by the late time decay of a scalar particle, which is radiatively induced by WIMPZILLA and the result is compared with the recent Planck 2013 data.  If the scalar particle is light enough, it also contributes to the invisible decay of the Higgs thus is constrained by the LHC data.

\vspace{0.5 cm}
\begin{acknowledgements}  This work is supported by Basic Science Research Program through the
National Research Foundation of Korea funded by the Ministry of
Education, Science and Technology (2011-0010294) and (2011-0029758).
\end{acknowledgements}

\end{document}